\documentclass{article}
\usepackage{amssymb}
\usepackage{amsmath}
\usepackage{graphicx}

\newcommand{\be}{\begin{equation}}
\newcommand{\ee}{\end{equation}}
\newcommand{\bea}{\begin{eqnarray}}
\newcommand{\eea}{\end{eqnarray}}

\newcommand{\D}{\partial}
\newcommand{\g}{\gamma}
\newcommand{\La}{\Lambda}
\newcommand{\e}{\varepsilon}

\newcommand{\p}{ {Poincar\'e} }
\newcommand{\ps}{ {Poincar\'e's} }

\begin{document}

\begin{titlepage}

\begin{centering}

{\huge {\bf Poincar\'e, the Dynamics of the Electron, and Relativity}}

\vspace{1.5cm}

\centerline{\large{Thibault Damour}} \vspace{.7cm}
\centerline{ {\it Institut
des Hautes Etudes Scientifiques}, } \centerline{{\it 35, Route de
Chartres,  F-91440 Bures-sur-Yvette, France} }

\vspace{5cm} \hrule

\begin{abstract}
On June 5, 1905 \p presented a Note to the Acad\'emie des Sciences entitled ``Sur la dynamique
de l' \'electron" (``On the Dynamics of the Electron"). 
After briefly recalling the context that led \p to write this Note, we comment its
content. We emphasize that \ps electron model consists in assuming
that the interior of the worldtube of the (hollow) electron  is filled with a {\it positive cosmological constant}.
We then discuss the several novel contributions to the physico-mathematical aspects of Special Relativity
which are sketched in the Note, though they are downplayed by \p who describes them as having
only completed the May 1904 results of Lorentz ``dans quelques points de d\'etail" (``in a few points of detail").
\end{abstract}
\hrule

\vspace{3mm}

\end{centering}

\vfill
\end{titlepage}

\section{Context}

In order to apprehend the meaning, and importance, of  Poincar\'e's Note, in the June 5, 1905 issue of the Comptes Rendus \cite{P05},
it is necessary to recall its context. [For  wider, and more detailed, historical studies of Poincar\'e's contributions to electrodynamics
see \cite{Darrigol1995,Darrigol1996,Darrigol2000,Darrigol2005,DarrigolBourbaphy,Galison,Walter,Weinstein2012}. 
For an access to Poincar\'e's works and archives see
\cite{Nancy}. See also the 2012 commemorative colloquium of the centenary of Poincar\'e's death at the Acad\'emie 
des Sciences \cite{Acad2012}.]

Poincar\'e gave sets of lectures on `` Electricit\'e et Optique ", at the Sorbonne,
in 1888, 1890 and 1899. These lectures were part of his duty as holder of a chair of ``Mathematical Physics and Probability Calculus"
(``Physique math\'ematique et calcul des probabilit\'es").
In his 1899 lectures (published as the last part of the book \cite{P01})
he expounded, in particular, Lorentz' approach to electrodynamics, which he considered as the most satisfactory one. 
Lorentz' approach had been developed a few years before, notably in Refs. \cite{Lorentz1892,Lorentz1895}. Lorentz' 1892 paper  \cite{Lorentz1892} already
implicitly contain the exact form (involving $1/\sqrt{1-v^2/c^2}$ factors) of the Lorentz transformation. See paragraph 138, pages 141-142
 in \cite{Lorentz1892} (one should only replace Lorentz' variable $t'$ by $t'_{\rm new}= \sqrt{1-v^2/c^2} \,  t'$ to get the exact
 Lorentz transformation). On the other hand, Lorentz's 1895 paper \cite{Lorentz1895} works most of the time only to first order in $v/c$,
 but expounds in clearer physical terms the usefulness of defining what Lorentz called there the ``local time", namely
 \be
 t' \equiv t - \frac1{c^2} {\bf v}\cdot \bar {\bf x} \,,
 \ee
 where
 \be
 \bar {\bf x} =  {\bf x} - {\bf v} t \,,
 \ee
 denotes the usual (Galilean-transformed) spatial coordinates in a moving frame. In addition, in paragraphs 89-92,
 Lorentz recalls his earlier (1892)  hypothesis (invented to explain the negative result of the
 Michelson-Morley experiment) according to which solid bodies moving with respect to the ether 
 get contracted (in the direction of the motion) by a factor $\simeq 1 - \frac12 \frac{v^2}{c^2}$.
 
 Both in his 1899 lectures \cite{P01}, and in his  invited
review talk on the ``Relations entre la physique exp\'erimentale et la physique math\'ematique"
 at  the ``Congr\`es International de Physique" taking place in Paris in 1900, \p
 expresses his dissatisfaction at Lorentz's approach, which is based on an accumulation of disconnected hypotheses 
 (famously referred to by \p as ``coups de pouce", i.e. ``nudges"; see citation below).
 In particular, he writes about the Lorentz(-Fitzgerald) contraction hypothesis that (p. 536 of \cite{P01}):
 
 `` Cette \'etrange propri\'et\'e semblerait un v\'eritable  ``coup de pouce" donn\'e par la nature pour \'eviter que le mouvement de la Terre puisse 
 \^etre r\'ev\'el\'e par des ph\'enom\`enes optiques. Ceci ne saurait me satisfaire et je crois devoir dire ici mon sentiment : je consid\`ere comme tr\`es probable que les ph\'enom\`enes optiques ne d\'ependent que des mouvements relatifs des corps en pr\'esence, sources lumineuses ou appareils optiques et {\it cela non pas aux quantit\'es pr\`es de l'ordre du carr\'e ou du cube de l'aberration, mais rigoureusement}. A mesure que les exp\'eriences deviendront plus exactes, ce principe sera v\'erifi\'e avec plus de pr\'ecision.
Faudra-t-il un nouveau {\it coup de pouce}, une hypoth\`ese nouvelle \`a chaque approximation ? Evidemment non : une th\'eorie bien faite devrait permettre de d\'emontrer le principe d'un seul coup dans toute sa rigueur. La th\'eorie de Lorentz ne le fait pas encore. De toutes celles qui ont \'et\'e propos\'ees, c'est elle qui est le plus pr\`es de le faire. On peut donc esp\'erer la rendre parfaitement satisfaisante sous ce rapport sans la modifier trop profond\'ement." [The italics are \ps.]
 
 Let us also mention that, in his  paper ``La th\'eorie de Lorentz et le principe de r\'eaction" \cite{P00}, written in 1900 at the occasion of the 25th anniversary of Lorentz's thesis, Poincar\'e discusses (as emphasized by O. Darrigol) the effect of an overall
translation, at some speed $v$, on the synchronization of clocks by the exchange of electromagnetic signals.
More precisely, he works only to {\it first order} in $v$, and notes that, if moving observers synchronize their watches by exchanging optical signals, and if they correct these signals by the transmission time under the (incorrect) assumption that the signals travel at the same speed in both directions, their watches will indicate not the ``real time'', but  the ``apparent time", say (denoting $\bar x \equiv x-vt$)
\be \tau = t - \frac{v \bar x}{c^2} + O(v^2).
\label{1} \ee
His main point is that the ``apparent time'' $\tau$ coincides with the formal mathematical variable
$t' \equiv t - \frac{v \bar x}{c^2} + O(v^2)$ introduced
by Lorentz in 1895 under the name of ``local time'' (and used by him to show the invariance of Maxwell's theory
under uniform translations, to first order in $v$).
 
 In addition, {\p}  mentions, in his 1902 book ``La science et l'hypoth\`ese" \cite{P02}, as one of the principles of physics,
  ``le principe du mouvement relatif" (see notably chapter VII), which he also refers to, at the end of chapter XIII, as ``le principe de relativit\'e"), and writes in chapter XIV
  (now attributing to Lorentz, what he was, in 1900, reproaching Lorentz not to take seriously enough),
  about the issue of whether experimental results might, one day, allow one to determine the absolute motion of the Earth:
  ``Lorentz ne l'a pas pens\'e ; il croit que cette d\'etermination sera toujours impossible ; l'instinct commun de tous les physiciens, les insucc\`es \'eprouv\'es jusqu'ici le lui garantissent suffisamment. Consid\'erons donc cette impossibilit\'e comme une loi g\'en\'erale de la nature ; admettons-la comme postulat. Quelles en seront les cons\'equences ? C'est ce qu'a cherch\'e Lorentz, $\cdots$."
  [In which one can particularly note the sentences, ``admettons-la comme postulat. Quelles en seront les cons\'equences ?", i.e. 
  ``let us admit the impossibility of detecting the absolute motion of the Earth as a postulate; and let us study the consequences of this
  postulate." Though he attributes this idea to Lorentz.]
  
On May 27, 1904, Lorentz publishes his breakthrough paper: ``Electromagnetic phenomena in a system moving with 
any velocity smaller than that of light" \cite{Lorentz1904}. [\p (probably informed by Lorentz) is soon aware of this paper.]
In the Introduction of his paper, Lorentz explicitly mentions, as a motivation for extending his previous 
results, the discontent expressed by \p  in his review talk  at  the 1900 ``Congr\`es International de Physique" in Paris.
In his 1904 paper, Lorentz defines some auxiliary variables, denoted  $(x',y',z',t')$, that are defined in terms of the 
 space and time coordinates $(x,y,z,t)$ measured in the ether frame, by the formulas
\bea \label{Lor}
x'&=& \g \ell \bar x\,, \nonumber \\
y'&=& \ell y \,, \nonumber \\
z'&=&\ell z \,, \nonumber \\
t'&=& \frac{\ell}{\g} t - \g \ell \frac{v}{c^2} \bar x \,.
\eea
Here, $\bar x$ (implicitly) denotes (as above) the Galilean-transformed $x$-coordinate, $\bar x \equiv x - vt$,
corresponding to a Galilean reference frame moving with the considered system, namely with the velocity $v$ in the $x$ direction.
In addition $\g$ denotes $\g \equiv 1/\sqrt{1-v^2/c^2}$ (which is actually denoted $\beta$ by Lorentz), while $\ell(|v|^2)$ denotes
an {\it a priori} undetermined rescaling factor, assumed to be some function of the squared modulus of the velocity $v$.
 
Lorentz shows that Maxwell's equations in vacuum are rigorously invariant under the change of variables \eqref{Lor}, provided the
electric and magnetic fields in the primed system are appropriately transformed. He also shows that the inhomogeneous Maxwell(-Lorentz)'s equations are approximately invariant when changing the charge and current densities by a transformation he writes down.

Then Lorentz makes two further assumptions: 

(A1) ``that the electrons, which I take to be spheres of radius $R$ in the state of rest,
have their dimensions changed by the effect of a translation, the dimensions in the direction of the motion becoming $\g \ell$ times,
and those in perpendicular directions $\ell$ times smaller"; and 

(A2) ``that the forces between uncharged particles, as well as those
between such particles and electrons, are influenced by a translation in quite the same way as the electric forces in an
electrostatic system".

Lorentz then computes the ``electromagnetic momentum" of a uniformly moving electron (assuming this accounts for the full
linear momentum of an electron, i.e. that the ``  `true', or `material' mass" of the electron vanishes) as being
\be \label{pLor}
{\bf p}^{\rm Lor} = \frac43 \frac{E_{\rm em}}{c^2} \g(v^2) \ell(v^2) \bf v ,
\ee
where 
\be \label{Eem}
E_{\rm em} = \int_{r>R} d^3 x \frac12 {\bf E}^2= \frac{e^2}{8\pi}\int_R^{\infty} \frac{dr}{r^2}=\frac{e^2}{8\pi R},
\ee
denotes the electrostatic energy of the field generated by a spherical, hollow electron, of radius $R$, at rest (the electric field is equal to $E=e/(4 \pi r^2)$
outside the electron, i.e. for $r>R$, and vanishes inside the spherical electron). [Like Lorentz and \p, we use here Heaviside units.] 
The factor $\frac43$ in \eqref{pLor} will be further commented below.

Requiring that the force law $d {\bf p}^{\rm Lor}/dt = \bf F$ [with ${\bf F}= e({\bf E} + \frac{\bf v}{c} \times {\bf B})$
being the (Lorentz) force] lead to consistent accelerations (using the transformation \eqref{Lor})  in the ether (rest) frame and in the moving frame, Lorentz derives the condition
\be
\frac{d \left(\g(v^2) \ell(v^2) v \right)}{d v} = \g^3(v^2) \ell(v^2) \,,
\ee
which implies
\be
\frac{d \left( \ell(v^2) \right)}{d v} =0,
\ee
and therefore [using $\ell(v^2) = 1 + O(v^2/c^2)$]
\be
\ell(v^2)=1.
\ee

With his definitions and his assumptions, Lorentz is then able to show the following theorem of ``corresponding states":
``If, in the system without translation, there is a state of motion in which, at a definite place, the components of [the polarization]
$\bf P$, $\bf E$ and $\bf B$ are certain functions of the time, then the same system after it has been put in motion
(and therefore deformed) can be the seat of a state of motion in which, at the corresponding place, the components of 
$\bf P'$, $\bf E'$ and $\bf B'$ are the same functions of the local time." In other words, there is an {\it active map} (given by 
Eqs. \eqref{Lor}, together with the field-transformation laws derived by Lorentz) between
a physical system at rest (in the ether) and a corresponding physical system moving with velocity $v$, such that the functions describing
the electromagnetic field generated by the system at rest, ${\bf E}(x,y,z,t)$ and ${\bf B}(x,y,z,t)$, are equal to
the corresponding transformed fields considered
as functions of the transformed variables, ${\bf E'}(x',y',z',t')$ and ${\bf B'}(x',y',z',t')$.
Note that Lorentz systematically emphasizes that $t'$ is not the ``true time", but rather the auxiliary ``local time" defined in
terms of the true time by the last equation \eqref{Lor}.

\p  (probably directly informed by Lorentz) appreciated the ``extreme importance" (as \p wrote to Lorentz in 1905)
of Lorentz's work and alluded to it (though not in detail) in his September 24, 1904 invited talk, on ``The principles of mathematical physics",
at the International congress of arts and science in Saint Louis (USA). His talk was delivered in French, and later translated \cite{PStLouis}.
Among the ``five or six\footnote{
This ambiguity comes from \ps 
last suggested general principle: ``I would add the principle of least action".}  general principles of physics", \p lists (in fourth position):

``The principle of relativity, according to which the laws of physical phenomena should be the same, whether for an observer fixed,
or for an observer carried along in a uniform movement of translation; so that we have not and could not have any means of discerning
whether or not we are carried along in such a motion." [This is the only principle which \p defines in detail.]

Note in this respect that I could not figure out whether \p was the first to use (in this sense) the expression ``principle of relativity" (which he had
already used, but only in passing, in his 1902 book ``Science and Hypothesis")\footnote{Lorentz writes in \cite{Lorentz1914}:
 ``Poincar\'e, on the contrary, obtained a perfect invariance of the equations of electrodynamics, and he formulated the `postulate of relativity', terms which he was the first to employ."}. 
By contrast, not only Lorentz does not use this
expression, but he does not seem to believe in the exact unobservability of a common translation. Indeed, he writes in his 1904 memoir \cite{Lorentz1904}:
``It would be more satisfactory if it were possible to show by means of certain fundamental assumptions and without neglecting terms of 
one order of magnitude or another, that many electromagnetic actions are entirely independent of the motion of the system."
[Note that Lorentz is not aiming at deriving an {\it exact} principle of relativity, but only a {\it partial} one (``many"). I think that he
had in mind the fact that the auxiliary time variable $t'$ differed (by a factor $1/\g$; see last equation \eqref{Lor}) from the ``true" time,
so that the unobservability was limited (as he says at the end of his paper) to non time-related experiments,
such as optical experiments ``in which the geometrical distribution of light and darkness is observed", or  in which ``intensities in adjacent
parts of the field of view are compared".]

Later in his talk, \p mentions that Lorentz's ``local time" is the (apparent) time indicated by the watches of two moving observers 
(``station $A$" and ``station $B$") when
they are synchronized by exchanging light signals and by (`wrongly' but conventionally)
assuming the isotropy of the speed of
light in the moving frame, {\it i.e.} the equality between the transmission times during the two exchanges
${A\to B}$ and ${B\to A}$. [However, he does not write down any equations, so that it is not clear
whether he is alluding to his previous {\it first order in $v$} result, (\ref{1}), or to an all order result.]
Finally, \p ends his Saint-Louis discussion of the principle of relativity (and of the related synchronization via light signals)
by asking the question:

``What would happen if one could communicate by non-luminous signals whose velocity of propagation differed from that of light ?
If, after having adjusted the watches by the optical procedure, one wished to verify the adjustment by the aid of these new signals,
then would appear divergences which would render evident the common translation of the two stations. And are such signals
inconceivable, if we admit with Laplace that universal gravitation is transmitted a million times more rapidly than light ?".

As we see, \ps reading of Lorentz's 1904 memoir had (in particular) induced \p to think about the connection between
``the principle of relativity" and gravitation. [See, however, also the reference to Langevin's Saint Louis talk below.]

\section{The main new results of \ps June 5, 1905 Note to the Comptes Rendus.}

In the Spring of 1905, \p started to study in detail Lorentz's memoir. His study led him to improve, and generalize, Lorentz's results. 
He announced some of his results in his June 5 Note \cite{P05}, reserving a detailed exposition to a long paper sent
for publication to the Rendiconti del Circolo Matematico di Palermo on July, 23 1905 \cite{P06}. This choice of medium of publication did not help
to publicize the novelty of \ps results. On the one hand, the Note, as we shall now discuss, is too short, too modest and too incomplete to convey
a clear idea of \ps achievements. On the other hand, the Palermo memoir is written in a rather obscure way, which hides some of the
most important new results of \p amidst very technical derivations. As a consequence, it seems that \ps achievements remained essentially
unnoticed until Minkowski studied them, extracted their essential core, and generalized them, in 1908 \cite{Minkowski1907,Minkowski1908,Minkowski1909}. [See, e.g. \cite{Walter1999,Damour2008} for  assessments of 
Minkowski's debt towards Poincar\'e.] Here, we shall limit ourselves to commenting the content of \ps June 1905 Note, emphasizing both its
importance, and its shortcomings.

\subsection{The first important result (according to \p himself): a dynamical derivation of the Lorentz-contraction of
moving electrons.}

The first point I wish to make concerns the title of \ps Note, namely ``Sur la dynamique de l'\'electron'' (``On the dynamics of the electron").
This title is quite different from the title used by Lorentz (``Electromagnetic phenomena in a system moving with 
any velocity smaller than that of light"). [It is also quite different from the title of  Einstein's paper on Relativity  \cite{Einstein1905} (``On the electrodynamics of moving bodies")\footnote{Note that Einstein's paper  was received by the {\it Annalen der Physik}  on June 30, 1905.}.]
Though \ps text does not make it so clear, I think that this title indicates that \p considers that his main new result consists
in ``dynamically deriving" one of the key assumptions of Lorentz, namely assumption (A1) above, stating (as an ad hoc hypothesis)
that a moving (hollow) spherical electron gets Lorentz-contracted into an ellipsoidal shape.
Let me also note that the only explicit article citation in \ps Note is Lorentz's 1904 memoir. 
[In addition,  \p cites the names of Michelson, Langevin, Kaufmann, Abraham and Laplace.]

\ps dynamical derivation of the contraction of each electron is initially based on a physical reasoning: because of the
 electrostatic self-repulsion, a hollow spherical electron of radius $R$ needs
 to be stabilized by some counteracting force, holding together the charge distribution on the shell of radius $R$.
 \p assumes that this counteracting force is ``{\it une sorte de pression constante ext\'erieure dont le travail est
 proportionnel aux variations du volume}" (`` {\it a kind of constant exterior pressure whose work is proportional to the variations
 of volume}"). [Note that this is one of the very rare sentences italicized by \p, thereby confirming its central role.]
 Alternatively, one can (as \p does) consider that there exist a {\it negative internal pressure} (i.e. an internal tension) which
 holds together the electron. \ps derivation is quite involved and is not even sketched in his Note. He only says that his derivation
 is based on ``an application of the principle of least action". His derivation was given in the long, follow-up Palermo article \cite{P06},
 submitted on July, 23 1905 (and published in January 1906). Let us sketch here the essence of \ps derivation, 
 anachronistically reformulated in modern notation, and terminology. 
 [We directly consider the case of interest leading to the Lorentz contraction, rather than to the electron models
 of Abraham or Langevin. See \cite{ElectromagneticMass} for more details and references on classical electron models.]
 
 \p assumes that the interior (labelled $\rm int$) of the electron worldtube is filled with a {\it positive cosmological constant} $\La$, corresponding
 to an action contribution (we use $c=1$ like \p, except when it is physically clarifying to use physical units)
 \be \label{SLa}
 S_\La= -\int_{\rm int} d^4 x  \, \La .
 \ee
 This action contribution is clearly Lorentz invariant.
 The total action for the electron then contains two terms:
 \be
  S_{\rm tot}=  S_{\rm em} + S_\La ,
 \ee
 where the first term, $S_{\rm em}$, is the electromagnetic action. When considered as a functional of both the electromagnetic 4-potential 
 $A_\mu$ and the sources (i.e. the 4-current $j^\mu$) the electromagnetic action reads
 \be
 - \frac14 \int d^4 x F^{\mu \nu} F_{\mu \nu}  + \int d^4 x \, A_\mu j^\mu \,,
 \ee
 where $F_{\mu \nu} = \D_\mu A_\nu -  \D_\nu A_\mu$. However, as emphasized in Ref. \cite{BraccoProvost2009}, \p  uses an action which is an (implicit) functional of the sources.
 The latter ``Fokker action" would generally be written (after replacing $A_\mu$ by its functional expression in terms of $j^\mu$,
 and using suitable integrations by parts) as $S_{\rm em}[j] = \frac12  \int d^4 x \, A_\mu j^\mu $. However, \p writes it in the
 equivalent (modulo integration by parts) form
 \be\label{Sem}
 S_{\rm em} =  + \frac14 \int d^4 x F^{\mu \nu} F_{\mu \nu} = \int dt d^3 x \frac12 \left( - {\bf E}^2 + {\bf B}^2 \right) \,,
 \ee
 where he (seemingly) considers that the electromagnetic field is expressed as a functional of the electric sources\footnote{However,
 \p does not specify the dynamics determining the 4-current.}. Note that the overall sign
 in \eqref{Sem} is the opposite of the usual field action $ S[A_\mu] =\int d^4x \frac12 \left(  {\bf E}^2 - {\bf B}^2 \right)$.
 [Somewhat confusingly, \p systematically works with  quantities that he calls ``actions", which have the opposite sign of the usual
 actions; so that he ends up with an electron ``action" equal to $+ m_{\rm electron} c^2 \sqrt{1 - \frac{{\bf v}^2}{c^2} }$ 
 instead of the standard, opposite definition, Eq.~\eqref{Lelectron} below.]
 
 Working with the standard sign, the total action for the electron used by \p reads
 \be \label{Stot}
 S_{\rm tot}= S_{\rm em}[j_\mu] + S_\La = + \frac14 \int d^4 x F^{\mu \nu}[j] F_{\mu \nu}[j]  -\int_{\rm int} d^4 x  \, \La  .
 \ee
 When considering the dynamics of a single electron, \p considers a  four-current $  j^\mu$ localized on the surface of the hollow electron worldtube,
 and reducing to a uniform electric charge density on a sphere of radius $R$ in the rest-frame of the electron, when the latter is in equilibrium. [When discussing an ellipsoidally deformed electron in \cite{P06}, \p assumes that it behaves as a conducting shell.]
 
 Given $\La$,  the value of the electron radius $R$ is determined (as emphasized by \p) by extremizing $ S_{\rm tot}$ with respect to $R$.
 In view of the formal relativistic invariance of $ S_{\rm tot}$, it is enough to work in the rest frame of the electron. The corresponding
 value of the total Lagrangian is clearly
  \be
 L_{\rm tot}(R)=- \frac12 \int d^3 x \, {\bf E}^2  - \La \frac{4\pi}{3} R^3 \,,
 \ee
where the electric field is ${\bf E} = e {\bf r}/(4 \pi r^3)$ for $r>R$, and vanishes for $r<R$.
This yields
 \be \label{L1}
  L_{\rm tot}(R)=  -  E_{\rm em}(R)  - \La \frac{4\pi}{3} R^3 \,,
 \ee
 where $  E_{\rm em}(R)$ denotes the electric field energy, 
 $  E_{\rm em}(R) = \frac{e^2}{8\pi R}$, as given in Eq. \eqref{Eem}. Extremizing $ L_{\rm tot}(R)$ with respect to $R$ then yields the condition
 \be \label{equilib}
 0= R \frac{ d L_{\rm tot}(R)}{d R}=  +  E_{\rm em}(R)  - \La 4\pi R^3 \,,
 \ee
 i.e. an equilibrium radius $R_*(\La)$ satisfying
 \be \label{equilib2}
 \La 4\pi R_*^4= \frac{e^2}{8\pi} \,.
 \ee
 We see that $\La$ must be positive. As the equilibrium value $R_*$ of $R$ corresponds to  maximizing $ L_{\rm tot}(R)$, i.e.
 minimizing the corresponding (rest-frame) Hamiltonian $ H_{\rm tot}(R)= - L_{\rm tot}(R)$, the \p electron model
 is stable under {\it spherical} perturbations. Surprisingly, \p  did not seem to worry about
 the more delicate issue of stability under non-spherical perturbations. 
 
 Following  the paragraph 6 of \ps Palermo article \cite{P06} (whose
 results and notation we follow, except that we do not use primes for rest-frame quantities),  it is easy to consider an ellipsoidally deformed electron (behaving like a hollow conductor) having, in its rest-frame, 
 a volume $\frac{4\pi}{3} {\bar R}^3$ and an ellipticity $\theta$ ($\theta = R_y/R_x=R_z/R_x$). The
 rest-frame Hamiltonian ($H_{\rm tot} =H'+F'$ in \ps notation)  depends on the 
 volume (i.e. on $\bar R$) and on the  ellipticity $\theta$ as 
 \be
H_{\rm tot}(\bar R, \theta)= - L_{\rm tot}(\bar R, \theta) = E_{\rm em}(\bar R) {\bar \varphi}(\theta) \theta^{2/3} + \La \frac{4\pi}{3} {\bar R}^3 \,,
 \ee
 where we denoted ${\bar \varphi}(\theta)=\varphi(\theta)/\varphi(1)$. The latter function is defined (from
 the Abraham electron model) by Eq. (5) in paragraph 6 of \cite{P06}, i.e. as the analytic continuation in $\theta$ (starting from
 the interval $0<\theta<1$) of
 \be
 {\bar \varphi}(\theta) = \left[\frac{1}{2 \e} \ln \frac{1+\e}{1-\e} \right]_{ \e=\sqrt{1-\theta^2} } \, .
 \ee

 One then finds that though the necessary condition of equilibrium of the ellipsoidally-deformed electron (namely that the function $H_{\rm tot}(\bar R, \theta)$
 have an {\it extremum} at $(\bar R,\theta)=(R_*,1)$) is satisfied, this equilibrium is unstable because the extremum of $H_{\rm tot}(\bar R, \theta)$
 is a {\it saddle point}: a local minimum with respect to the $\bar R$ axis, but
 a local maximum with respect to the $\theta$ axis\footnote{The constraint $\left[\varphi'(\theta)/\varphi(\theta)\right]_{\theta=1}=-2/3$
 derived by \p is equivalent to extremizing $\psi(\theta) \equiv {\bar \varphi}(\theta) \theta^{2/3}$ at $\theta=1$. However, this is a local maximum.}. This shows the {\it instability} of \ps electron model with respect to
 shape variations (under his Abraham-like assumption that
 the electric charge is distributed on the surface of the electron as if it were a conductor). Lorentz, who had worried about the stability of \ps
 electron under shape variations, had also found its instability under the assumption that the electric 
 charge is uniformly distributed, keeping fixed the total area (see Note {\bf 80} in \cite{Lorentz1916}). This ellipsoidal instability problem
of the  \p electron model is shared by the Bucherer-Langevin model \cite{Ehrenfest1906}, as well as by the Dirac ``extensible model of the electron" \cite{Gnadig1978}, in which \ps  volumic interior  negative pressure is replaced by the surface tension 
 of an elastic, charged conducting membrane (with vanishing electromagnetic field inside) \cite{Dirac1962}.
 
 From Eq. \eqref{equilib}, \p (implicitly) deduces that the last ($\La$-related) contribution to the electron Lagrangian \eqref{L1}
 is equal to $-\frac13 E_{\rm em}(R)$, so that, after extremization on $R$, the value of the electron Lagrangian in its
 rest frame is $- E_{\rm em}(R_*) -\frac13 E_{\rm em}(R_*) = - \frac43  E_{\rm em}(R_*)$.
 
 Finally, by an analysis equivalent to using the relativistic covariance of the action \eqref{Stot}, \p deduces that the Lagrangian describing,
 in the ether frame, the dynamics of electrons moving in a quasi-stationary manner is
 \be \label{Lelectron}
 L_{\rm electron} = - m_{\rm electron} c^2 \sqrt{1 - \frac{{\bf v}^2}{c^2} } \,,
 \ee
 where
 \be \label{melectron}
m_{\rm electron} = \frac43 \frac{E_{\rm em}}{c^2} \,.
\ee

 An alternative way (not available to \p) of getting these results is to combine Einstein's famous result of September 1905,
 namely
 \be
 m = \left[ \frac{E_{\rm tot}}{c^2}\right]_{\rm rest \, frame}= \frac1{c^2} \left[ \int d^3 T^{00}_{\rm tot}\right]_{\rm rest \, frame} \,,
 \ee
 with von Laue's well-known (virial) theorem stating that
 $\int d^3 T_{\rm tot}^{ij}=0$ for a body in a stationary state (in its rest frame) \cite{Laue1911}. 
 
 The total stress-energy tensor
 corresponding to the action \eqref{Stot} is [we use the signature $\eta_{\mu \nu}= {\rm diag} (-1,+1,+1,+1)$]
 \be
 T_{\rm tot}^{\mu \nu}=  T_{\rm em}^{\mu \nu} + T_{\La}^{\mu \nu} =T_{\rm em}^{\mu \nu} - \La \eta^{\mu \nu} \theta(R-r) ,
 \ee
 where $\theta(x)$ denotes Heaviside's step function.
[Note in passing that the pressure corresponding to $T_{\La,  {\rm int}}^{\mu \nu}=  - \La \eta^{\mu \nu}$ is $- \La <0$,
as emphasized by \p.]  Using $\eta_{\mu \nu} T_{\rm em}^{\mu \nu} =0$ and $\int d^3 T_{\rm tot}^{ii}=0$ immediately yields
$\int d^3 x T_{\rm em}^{00} = - \int d^3 x T_{\La}^{i i}$, i.e.
\be
E_{\rm em} = \int d^3 x T_{\rm em}^{00}= 3 \int_{r<R} d^3x \La = 3 E_\La \,,
\ee
in agreement with Eq. \eqref{equilib}.

Summarizing \ps result in modern terms: the condition of dynamical equilibrium between the electrostatic self-repulsion and the inner
(negative) pressure $- \La$ associated with the additional action \eqref{SLa} yields an additional (positive)
energy contribution $E_\La = \int_{r<R} d^3x \La $ related to the electromagnetic rest energy $E_{\rm em} = \int d^3 x T_{\rm em}^{00}$
by the simple relation
\be
E_\La =\frac13 E_{\rm em}.
\ee
As a consequence, the total rest energy of the electron is
\be
E_{\rm tot} = E_{\rm em} + E_\La =\frac43 E_{\rm em} \,,
\ee
which corresponds to the electron mass \eqref{melectron} (implicitly) derived by \p.

Alas, though it is likely that \p had derived Eqs. \eqref{Lelectron}, \eqref{melectron} when he wrote his short Note,
he did not display these results then, but buried them, in the middle of a rather abstruse technical discussion, in his
47-pages long Palermo memoir.

\subsection{The second important result (according to \p himself): defining the class of relativistically-invariant gravitational force laws.}

\p writes that it is important to examine in more detail the second assumption of Lorentz, namely assumption (A2) above
``that the forces between uncharged particles, as well as those
between such particles and electrons, are influenced by a translation in quite the same way as the electric forces in an
electrostatic system". This led him to define, for the first time, the class of relativistically-invariant, action-at-a-distance gravitational force laws,
between two (arbitrarily moving) masses.

However, in his June 5 Note, he barely sketches his results, only saying that: (i) he succeeded in constructing such Lorentz-invariant force laws;
(ii) he assumed that the ``propagation of gravitation is not instantaneous, but takes place at the velocity of light"  (which makes
\p speak of a ``{\it gravitational wave}" (``{\it onde gravifique}") leaving the attracting body and propagating towards the attracted one);
and (iii) he could arrange his construction so that the fractional deviations from Newton's non-relativistic $1/r^2$ law were only
of order $v^2/c^2$.

Again, he buried the most interesting, and most novel aspects of his results, in the last pages of his Palermo article. 
Indeed, it is only in his Palermo memoir that \p remarks (in the middle of some technical developments) that:

(R1) the Lorentz transformations leave invariant the quadratic form (with $c=1$) $x^2+y^2+z^2-t^2$; and,

(R2) the Lorentz transformations can be viewed as ``rotations" in ``4-dimensional space"
with coordinates `` $x, y, z, t \sqrt{-1}$". [The latter remark is used by \p to find all the relativistic invariants
constructible with what we would call today the two spacetime points of the attracted and attracting bodies, $x^{\mu}$, 
$x^{\mu}_1$, and their 4-velocities $d x^{\mu}/ds$, $dx^{\mu}_1/ds_1$.]

Actually, I find rather likely that \p had not fully understood the latter results (especially the ``Wick rotation" (R2)) 
at the time when he was writing his Note.
Indeed, the results (R1) and (R2) just quoted do not enter the Palermo article in the first paragraphs (which display most
of the results alluded to in the Note), but only (somehow in passing) at the end of paragraph 4 (for the remark (R1)),
and in the middle of paragraph 9 (for the remark (R2)). When reading \p (and especially his Palermo article), I have
the impression that \p had not well planned his article, but was finding new results as he proceeded with his technical developments,
and was incorporating them without coming back and rewriting the first paragraphs in a more logical manner.

\subsection{Other important results (as viewed today).}
 
 The results discussed above are, in my opinion, the ones that \p himself would have recognized as significant new contributions.
 \p was known to be very critical, and demanding, when presented with supposedly new scientific results. His usual reaction was
 to say: ``A quoi bon ?" (``What is it good for ?"). He applied this demanding criterion to himself. This is why many other
 results mentioned in his short Note (which we appreciate today as having broken new ground in Relativity) are qualified by \p
 himself as being only incremental additions to Lorentz's 1904 memoir. Indeed, at the beginning of his Note, just before
 starting to write his first equation, \p writes:
 
 ``les r\'esultats que j' ai obtenus sont d'accord sur tous les points importants avec ceux de Lorentz; j'ai \'et\'e seulement conduit
 \`a les modifier et \`a les compl\'eter dans quelques points de d\'etail."
 
 Let me, however, indicate some of these ``points of detail" where \p actually went significantly beyond Lorentz.
 
 First, and foremost, there is the fact that \p begins his Note by asserting the principle of relativity as being exact: 
 
 ``Il semble que cette impossibilit\'e de d\'emontrer le mouvement absolu soit une loi g\'en\'erale de la nature."
 
Then \ps first equation (which he calls the ``Lorentz transformation") differs from Lorentz' original
 writing, Eqs. \eqref{Lor} (though it is equivalent to it). \p writes
 \bea \label{LorP}
x'&=& k \, \ell \, (x + \e t) \,, \nonumber \\
y'&=& \ell \,y \,, \nonumber \\
z'&=& \ell \,z \,, \nonumber \\
t'&=&  k \,\ell \,( t  + \e x) \,,
\eea
with
\be
k\equiv\frac1{\sqrt{1-\e^2}}.
\ee
Curiously enough, \p never says that $\e =- v/c$, but only says that ``$\e$ is a constant which defines the transformation".
He also neglects the physical difference (important for Lorentz) between the ether-frame coordinates, and the transformed ones,
just saying: ``$x, y, z$ are the coordinates and $t$ the time before the transformation, $x', y', z'$ and $t'$ after the transformation".
These are two examples of his (fruitful) mathematical approach to the problem (which, for example, makes natural for him to define
the ``velocity after the transformation" as $ d x'/dt', dy'/dt', dz'/dt'$, while Lorentz was always working with the
Galilean-transformed velocity $v'^x=(dx/dt) -v$, $v'^y=dy/dt$, $v'^z=dz/dt $).

The first new result of \p is to say that the transformations \eqref{LorP} (together with spatial rotations) ``must form a group",
and that this requires the unknown function $\ell(v^2)$ to be equal to 1. He adds that ``this is a consequence that Lorentz
had obtained by another route." 

Actually, it seems to me that, here, \p is (fruitfully) betraying Lorentz's approach. Indeed, 
for Lorentz, the set of Lorentz transformations does not need to form a group, because these transformations must
always transform the ether coordinates (modulo spatial rotations and time shifts) into some auxiliary variables attached
to a moving frame parametrized by $\bf v$. It makes no sense (for Lorentz) to compose two Lorentz transformations, 
$(x,y,z,t) \to (x',y',z',t') \to (x'',y'',z'',t'')$ 
because the only physically meaningful transformations are $(x,y,z,t) \to (x',y',z',t')$ and $(x,y,z,t) \to (x'',y'',z'',t'')$.
We have here an example of a (fruitful) contradiction between \p the mathematician, and \p the physicist. If asked,
\p the physicist would have agreed with Lorentz that only $t$ (in the ether frame) measures the ``real time", while
$t'$ measures some (fictitious) ``ideal time" (see below). On the other hand, \p the mathematician considers all
the different variables $(x,y,z,t); (x',y',z',t'); (x'',y'',z'',t'')$ as being on an equal footing, which allows him to consider
the group composition $(x,y,z,t) \to (x',y',z',t') \to (x'',y'',z'',t'')$. [By contrast, it was natural for Einstein the physicist 
to compose transformations, because he was really considering all inertial coordinates as being physically on the
same footing \cite{Einstein1905}.]

Note in passing that, though \p used in his derivations the relativistic law of composition of velocities, both for composing
a boost velocity with the electron velocity, and for composing two boost velocities, e.g.
\be \label{v+v}
\e'' = \frac{\e+\e'}{1+ \e \e'} \,,
\ee
 he does not bother to mention it
in his Note. [It will, however, be explicitly used several times in his Palermo memoir.]

The second new result of \p (which actually corrects a very important shortcoming of Lorentz's paper) is to give,
for the first time, the correct transformation law for the electromagnetic four-current $j^{\mu}$. In a modern notation,
this law reads
 \bea \label{LorjP}
j'^x&=& \frac{k}{\ell^3}\, (j^x + \e j^t) \,, \nonumber \\
j'^y&=& \frac{1}{\ell^3}\,j^y \,, \nonumber \\
j'^z&=& \frac{1}{\ell^3} j^z\,, \nonumber \\
j'^t&=&   \frac{k}{\ell^3}\,( j^t  + \e j^x) \,,
\eea
and \p certainly understood that $(j^x,j^y,j^z,j^t)$ varied, under a Lorentz transformation (when $\ell=1$), like $(x,y,z,t)$ [i.e. that they are
both 4-vectors]. Armed with his definition \eqref{LorjP}, \p showed (for the first time) the (exact) relativistic invariance of the
inhomogeneous Maxwell-Lorentz equations.

Let me also mention that, in his Palermo article, \p obtains further important results on the mathematics of Relativity. Notably,
he computes the Lie algebra of the Lorentz group, showing in particular that the commutator of two boosts is a rotation:
\be \label{commutator}
\left[ t \D_x + x \D_t,  t \D_y + y \D_t \right] = x \D_y - y \D_x \,.
\ee
He understands and uses the fact that (in the Lorenz gauge) the 4-potential $A^\mu$ transforms as $x^\mu$.
In addition, he discusses many invariants of the (\p and) Lorentz group, such as the ``Minkowski" scalar product of two four-vectors,
and the electromagnetic invariants $ {\bf E}^2 - {\bf B}^2$ and ${\bf E} \cdot{\bf B}$.
 
\section{Conclusions}

In conclusion, \ps June 5, 1905 Note announces important mathematical and physical advances in what we would call today,
Special Relativity, relativistic electrodynamics and relativistic gravitation (see the partial summary above). 
One can, however, regret that this Note did not explicitly
report some of the most important advances made by \p, and only published in the follow-up Palermo memoir, such
as: 

(1) the relativistic law of addition of velocities \eqref{v+v}; 

(2) the relativistic electron Lagrangian \eqref{Lelectron}, \eqref{melectron};

(3) the understanding of Lorentz transformations as  ``rotations" in a ``4-dimensional space"
with coordinates `` $x, y, z, t \sqrt{-1}$", and the associated method of constructing relativistic invariants ; 

(4) the Lie algebra of the Lorentz group \eqref{commutator}; and

(5) a more explicit description of the class of
relativistically-invariant, action-at-a-distance gravitational force laws, between two (arbitrarily moving) masses, and 
of the associated, inter-body ``gravitational wave" propagation effects.

My feeling is that \p wrote his Note too soon, before starting the writing of his long Palermo memoir, during which he finalized, as he
proceeded, some of his most important advances [such as (3) and (4) above]. If \p had written, say  in July 1905, a  Note 
containing a more complete summary of his Palermo article, with technical indications of the results (1)-(4) above, this Note
might have been included in the booklet on the Principle of Relativity, edited by Sommerfeld, collecting the important original 
papers on Relativity \cite{PrincipleRelativity}. One can understand why Sommerfeld decided  to include in his collection 
neither the June Note (whose relativistic content is too slim), nor the Palermo memoir (from which one cannot easily extract a self-contained 
account of its scattered relativistic content), but resorted instead to adding some notes after the Minkowski Cologne lecture \cite{Minkowski1908}
in which he mentioned several (though not all\footnote{In particular, Sommerfeld does not mention \ps use of a ``4-dimensional space"
with coordinates `` $x, y, z, t \sqrt{-1}$"}) of \ps relevant achievements.

It seems that neither \ps Note, nor his follow-up Palermo article, attracted much attention, at the time. The earliest citation of \ps
Note that I found is in the March 2, 1906 paper by  the experimental physicist (of electron-dynamics fame) Walter Kaufmann \cite{Kaufmann1906}.
What attracted Kaumann's attention was \ps introduction of a pressure to account for the internal
constitution of the electron. Concerning the dynamics of electrons, he speaks of the ``Lorentz-Einstein basic assumption".

On March 8, 1906, Lorentz belatedly acknowledges the reception of the Palermo memoir, which he claims to 
``have studied with the greatest interest"
(``Inutile de vous dire que je l'ai \'etudi\'e avec le plus grand int\'er\^et, et que j'ai \'et\'e tr\`es heureux de voir mes conclusions confirm\'ees
par vos consid\'erations."). However, he does not seem to have taken a full measure of the new contributions brought by Poincar\'e, some of which
were already mentioned in his June 1905 Note. Indeed, in the set of lectures he gave in March and April 1906 in Columbia University, New York, 
Lorentz \cite{Lorentz1916} (who cites the Palermo memoir but not the June 1905 Note) only credits \p for having introduced
 a constant ``normal stress $S$" that ``makes much clearer" how 
a spherical electron at rest can Lorentz-contract and remain in equilibrium when moving at any velocity $v<c$. 
Though he borrowed  \ps notation $k$ for the Lorentz ``$\g$ factor", and uses the ``modern" writing \eqref{LorP}
for the Lorentz transformation, Lorentz attributes only to Einstein several of the results already displayed in the June 1905 Note,
notably the  correct transformation law \eqref{LorjP} for the electromagnetic four-current $j^{\mu}$.

It would be interesting to study who else cited \ps Note in the early years of Special Relativity. 
But it seems clear that \ps short and cryptic  Note \cite{P05}, and its long and opaquely written follow-up paper \cite{P06}
(published in January 1906 in a mathematics journal), had difficulties
competing with the very transparently written, conceptually novel, quasi-simultaneous paper of Einstein \cite{Einstein1905} (published 
on September 26, 1905 in a very well-read physics journal).

I emphasized above how \p seems to downplay the novelty of some of his own results. \p was certainly over-modest in repeatedly stating that
he was only bringing incremental additions to Lorentz's results. On the other hand,
when focussing one's attention on \ps contribution (as we did above), without studying in detail the state of the art of the field
around 1905, one runs the risk of over-appreciating some of \ps results. For instance, while preparing this text, I had
a brief look at some of the contemporary articles of Langevin, notably his lecture \cite{Langevin1906} at the 1904 Saint-Louis international 
conference\footnote{Both \p and Langevin gave invited talks there; moreover, they spent a week together ``in the vast plains
of North America", on their way back from Saint Louis, which gave them ample time to discuss \cite{Langevin1913}.}, and a Note to the Comptes Rendus published by Langevin in 1905 \cite{Langevin1905}, 
just before \ps Note.

Langevin's Saint-Louis talk \cite{Langevin1906} reviews, in particular, Lorentz's 1904 results and mentions the issue of whether gravitation might
not transform as assumed by Lorentz and thereby allow one to observe one's motion with respect to the ether. In addition, Langevin
emphasizes the problem of finding which extra forces might ensure the stability of the electron:
``It seems necessary to admit something else in its structure than its electric charge, an action which maintains the unity of the electron and prevents its charge from being dissipated by the mutual repulsions of the elements which constitute it." He even mentions 
the possibility that gravitation ``acts at the interior of the electrons in order to insure their stability", though he is aware
that gravity is much weaker than electromagnetism. [Let me mention in passing that Langevin also writes: 
``A very simple calculation shows also that the stock of energy represented by the electric and magnetic fields surrounding the electrons contained in an atom is sufficiently great to supply for ten million years the evolution of heat discovered by Curie in the radium salts."
In view of the then classic result \eqref{melectron} we see that Langevin probably had in mind the pre-Einsteinian energy $E_{\rm em}=\frac34 m_{\rm electron} c^2$.]

 In addition, Langevin's 1905 Note \cite{Langevin1905} (which is entitled
 ``On the physical impossibility to observe [mettre en \'evidence] the translatory motion of the Earth") reports the result that the Lagrangian of any electrified system transforms under a Lorentz boost as $L'=L \sqrt{1 - \frac{{\bf v}^2}{c^2} }$.
This might explain why \p felt that his result \eqref{Lelectron} might not be a novelty [in addition, \p certainly knew that Lorentz's result
\eqref{pLor} implied the Lagrangian \eqref{Lelectron} (modulo an additive constant)].

It is probable that the first person to fully grasp the importance of \ps results concerning the mathematical aspects 
of Special Relativity, and relativistic gravitation, was Minkowski. In his first (November 1907) work \cite{Minkowski1907}
(which is a digest of \ps results), Minkowski cites six times Poincar\'e's name (and only twice Einstein's name) [see \cite{Damour2008}
for a detailed discussion.] 
On the other hand, Minkowski  decided {\it not to cite at all}
\p when he delivered (and then wrote up) his famous September 1908 Cologne lecture on  ``Raum und Zeit" (``Space and Time").
[For discussions of Minkowski's attitude towards \p see, e.g., \cite{Walter1999,Damour2008}.] As already mentioned, Sommerfeld
completed the reprinting of Minkowski's article in his booklet \cite{PrincipleRelativity} by mentioning some of \ps results. 
It seems that  people  fully recognized \ps contributions to Relativity only after his death, see notably the
scientific obituaries of Langevin \cite{Langevin1913} and Lorentz \cite{Lorentz1914}. Let me also mention the fair, and complete,
account of \ps contributions to Relativity  given by the young Pauli (prompted by the old Klein who insisted on citing the
contributions of \p) in his 1921 Encyclopedia article \cite{Pauli1921}. 

To end this account, let me recall the well-known fact that, in spite of \ps repeated pleads (especially in his non-technical
lectures) for the importance of the ``Principle of Relativity", in spite of his (pre-Einstein) discussions of clock synchronization by
means of electromagnetic signals, and in spite of his important technical results concerning
the mathematical aspects of Relativity, it seems that he never abandoned his conviction that there exists an absolute
time and an absolute space [see, e.g., the discussion around Eq. (1.6) in \cite{Damour:2005pe}]. He also, apparently, never cited, nor probably appreciated, Einstein's contributions to Relativity.
Indeed, a few months before his death, in a lecture given on May 4, 1912 at the University of London \cite{P12},
after recalling what we would call today the ``Poincar\'e-Minkowski" picture
of ``time as a fourth dimension of space", with rotations among  `` $x, y, z, t \sqrt{-1}$", etc. he concludes his
lecture by pleading for ``keeping one's old habits":  

``Quelle va \^etre notre position en face de ces nouvelles conceptions? Allons-nous \^etre forc\'es de modifier nos conclusions? 
Non certes : nous avions adopt\'e une convention parce qu'elle nous semblait commode, et nous disions que rien ne pourrait nous contraindre \`a l'abandonner. AujourdÕhui certains physiciens veulent adopter une convention nouvelle. Ce n'est pas qu'ils y soient contraints; ils jugent cette convention nouvelle plus commode, voil\`a tout ; et ceux qui ne sont pas de cet avis peuvent l\'egitimement conserver l'ancienne pour ne pas troubler leurs vieilles habitudes. Je crois, entre nous, que c'est ce qu'ils feront encore longtemps."

Moreover, in his last lectures (July 1912) ``on the dynamics of the electron" \cite{P13}, he insists on saying that the Lorentz transformation maps
`` a {\it real} phenomenon which takes place in $x, y, z$, at the instant $t$, into an 
{\it ideal} phenomenon
which is its image, and which takes place in $x', y', z'$, at the instant $t'$" (with italics added by me on the words ``r\'eel" and ``id\'eal").
By contrast, let me recall that Einstein once said that his main new insight was to realize that the auxiliary time variable $t'$ used by
Lorentz was ``time, pure and simple".

Let us finally mention that Lorentz (in spite of his high appreciation of Einstein), also always kept his ``old habits" about absolute space
and absolute time. One has an insight on how Lorentz (and probably also \p) thought about the issue of the ``principle of relativity" through
Lorentz's comments (in his book \cite{Lorentz1916}) on ``the many highly interesting applications that Einstein made of this principle [of Relativity]". In particular, Lorentz writes (p. 230):

``[...] the chief difference being that Einstein simply postulates what we have deduced, with some difficulty and not altogether satisfactorily, from the fundamental equations of the electromagnetic field."
 
 It is probable that \p also felt that Einstein was simply postulating what had to be proven from underlying microscopic dynamical considerations.
 I think that when \p was using the word ``principle" he had (mostly) in mind a general physical property that is rooted in (and provable from)
  some microscopic dynamics (see, e.g., the citation from \cite{P01} in the first section above : 
  ``une th\'eorie bien faite devrait permettre de d\'emontrer le principe d'un seul coup dans toute sa rigueur."); 
  by contrast, Einstein used his Principle of Relativity as a primitive symmetry requirement restricting the
  laws of physics.

\section*{Acknowledgments} 
 I thank Olivier Darrigol for informative email exchanges.

\end{document}